\documentclass [11pt,a4paper] {article}

\usepackage[cp1252]{inputenc}

\usepackage{amssymb}
\usepackage{amsmath}
\usepackage{amsfonts,amssymb}

\usepackage[dvips]{graphicx}

\usepackage{bbm}

\DeclareMathAlphabet{\mathpzc}{OT1}{pzc}{m}{it}

\begin{document}

\title{Contextuality and truth-value assignment}

\author{Arkady Bolotin\footnote{$Email: arkadyv@bgu.ac.il$\vspace{5pt}} \\ \textit{Ben-Gurion University of the Negev, Beersheba (Israel)}}

\maketitle

\begin{abstract}\noindent In the paper, the question whether truth values can be assigned to the propositions before their verification is discussed. To answer this question, a notion of a propositionally noncontextual theory is introduced that in order to explain the verification outcomes provides a map linking each element of a complete lattice identified with a proposition to a truth value. The paper demonstrates that no model obeying such a theory and at the same time the principle of bivalence can be consistent with the occurrence of a non-vanishing ``two-path'' quantum interference term and the quantum collapse postulate.\\

\end{abstract}

\section{Introduction}\label{Introduction}  

\noindent As it is known, the presence of contextuality in quantum theory makes it impossible to view a measurement as merely revealing pre-existing properties of a quantum system \cite{Kochen}. More specifically, no hidden-variable model in quantum theory can assign $\{0,1\}$-valued outcomes to the projections of the measurement in a way that depends only on the projection and not the context in which it appeared, even though the Born probabilities associated with those projections are independent of the context.
$\,$\footnote{\label{f1}For an overview of different aspects of contextuality in quantum theory and beyond, see \cite{Liang,Dzhafarov}. Also, see a review of the framework of ontological models in \cite{Simmons,Harrigan}.\vspace{5pt}}\\

\noindent On the other hand, thanks to a trivial probability function mapping true propositions to probability 1 and false propositions to probability 0, after the assertion of the truth of propositions the sum of the Born probabilities can be presented as a disjunction of a set of the propositions where exactly one proposition is true while the others are false. This naturally raises the question whether one can assign truth values to propositions about properties of a state of a quantum system \textit{before the act of verification}. In more general terms, can the assignment of pre-existing truth values to all the lattice elements associated with the system under investigation be always made available in quantum theory?\\

\noindent In this paper, we introduce a notion of a propositionally noncontextual theory, which can provide a map linking each element of a complete lattice to a truth value so as to explain the verification outcomes of experimental propositions associated with the state of the system. Using a quantum version of the double-slit experiment as an example, this paper demonstrates that no model based on such a theory and at the same time obeying the principle of bivalence can agree with the occurrence of a non-vanishing ``two-path'' quantum interference term and the quantum collapse postulate.\\

\section{Pre-existing truth values and the principle of bivalence}\label{Two}  

\noindent Let us consider a double-slit quantum interference set-up in which detectors placed just behind slits indicate by a record a particle passage through a particular slit. Let $X_1$ denote the proposition of a click of the detector behind slit 1 such that $X_1$ is true (``1'') if the detector 1 clicks (verifying in this way that the particle has indeed passed through slit 1) and $X_1$ is false (``0'') if this detector does not click (thus verifying that the particle has in fact not passed through slit 1). Let $X_2$ analogously denote the proposition of the second detector's click.\\

\noindent Let us introduce the proposition:\smallskip

\begin{equation} \label{1}  
      X_{12}
      \equiv
      X_1 \,\underline{\lor}\, X_2
      \equiv
      (X_1 \lor X_2) \!\land\! \neg (X_1 \land X_2)
    \;\;\;\;  ,
\end{equation}
\smallskip

\noindent where the symbol $\underline{\lor}$ stands for the associative and commutative operation of \textit{exclusive disjunction} that outputs true when one of its inputs is true and the other is false. This proposition corresponds to the assertion that the particle passes through exactly one slit -- either 1 or 2. Subsequent to the recording of which-slit information (i.e., after the detectors confirms the particle's passage through either slit), the proposition $X_{12}$ represents an exact (i.e., sharp) property that the combined system – i.e., the particle plus the detectors – possesses.
$\,$\footnote{\label{f2}For an approach to unsharp (and partial) forms of a quantum logic, see \cite{Chiara}.\vspace{5pt}}\\

\noindent To keep things general, let us consider a complete lattice $\mathcal{L}=(L,\sqcup,\sqcap)$ containing any set $L$ where each two-element subset $\{y,z\} \subseteq L$ has a join (i.e., a least upper bound) and a meet (i.e., a greatest lower bound) defined by $y \sqcup z \equiv l.u.b.(y,z)$ and $y\sqcap z \equiv g.l.b.(y,z)$, correspondingly.\\

\noindent In addition to the binary operations $\sqcup$ and $\sqcap$, let the lattice $\mathcal{L}$ contain a unary operation $\sim$ defined in a manner that $L$ is closed under this operation and $\sim$ is an involution, explicitly, ${\sim}y \in L$ if $y \in L$ and ${\sim}({\sim}y)=y$.\\

\noindent Let $\alpha(Y) = {[\![ Y ]\!]}_v$, where $Y$ is any proposition associated with an exact property of the system, refer to a valuation, i.e., a mapping $\alpha: S \to \mathcal{V}_N$ from a set of propositions $S=\{Y\}$ to a set $\mathcal{V}_N = \{ \mathfrak{v} \}$ of truth-values $\mathfrak{v}$ ranging from 0 to 1 where $N$ is the cardinality of $\{ \mathfrak{v} \}$.\\

\noindent Let us introduce the following definition: Suppose that there is a homomorphism $f: L \to S$ and let $y \in L$ be a lattice element identified with the proposition $Y \in S$. Then, a theory will be defined as \textit{propositionally noncontextual} if to explain (predict) the verification outcomes (i.e., the truth values of the proposition $Y$) it provides a truth-function $v$ that maps each lattice element to the truth value of the corresponding proposition, namely, $v(y)={[\![ Y ]\!]}_v$, basing on the following principles: $v(y)=0$ if $y=0_L$ and $v(y)=1$ if $y=1_L$, where $0_L$ is the least and $1_L$ is the greatest elements of the lattice (identified with always true and always false propositions, respectively).\\

\noindent Correspondingly, a theory, which is not propositionally noncontextual, will be defined as propositionally contextual.
$\,$\footnote{\label{f3}This definition is motivated by a similar one introduced in the paper \cite{Arora}.\vspace{5pt}}  \\

\noindent Then, allowing that within a propositionally noncontextual theory the following valuational axioms hold $v(y \sqcup z)={[\![ Y\lor Z ]\!]}_v$, $v(y \sqcap z)={[\![ Y\land Z ]\!]}_v$, $v(\sim y)={[\![ \neg Y ]\!]}_v$, the truth value of the compound proposition $X_{12}$ can be expressed in the lattice-theoretic terms as follows\smallskip

\begin{equation} \label{2} 
    v(x_{12})
    =
   v
   \left(
      \left({\sqcup}_{i=1}^{2} x_i \right)
       \!\sqcap\!
      \left({\sim}\left({\sqcap}_{i=1}^{2} x_i \right)\right)
   \right)
    =
   {[\![ X_{12} ]\!]}_v
   \;\;\;\;  ,
\end{equation}
\smallskip

\noindent where the lattice elements $x_i$ are attributed to the propositions $X_i$ such that $v(x_i)=[\![X_i]\!]_v$.\\

\noindent Alternatively, the truth values of logical connectives \textit{disjunction}, \textit{conjunction} and \textit{negation} can be decided through the corresponding truth degree functions ${[\![ Y\lor Z ]\!]}_v = F_{\lor} ({[\![ Y ]\!]}_v, {[\![ Z ]\!]}_v)$, ${[\![ Y\land Z ]\!]}_v = F_{\land} ({[\![ Y ]\!]}_v, {[\![ Z ]\!]}_v)$, and ${[\![ \neg Y ]\!]}_v = F_{\neg} ({[\![ Y ]\!]}_v)$. For example, in {\L}ukasiewicz logics, the definition of a truth degree function of a negation connective $F_{\neg}$ is $1-{[\![ Y ]\!]}_v$. To meet the {\L}ukasiewicz version of negation, we will accept that\smallskip

\begin{equation} \label{3} 
    v({\sim}y)
    =
   1 - v(y)
   \;\;\;\;  .
\end{equation}
\smallskip

\noindent In accordance with this definition, $\sim 0_L=1_L$ and $\sim 1_L=0_L$ meaning that the lattice greatest element and the lattice least element are complements of each other.\\

\noindent Furthermore, as it is stated in \cite{Pykacz17}, {\L}ukasiewicz versions of disjunction and conjunction coincide with the truth-functions of the lattice joins and meets, namely, $v(y \sqcup z)=\min{\{v(y)+v(z),1\}}$ and $v(y \sqcap z)=\max{\{v(y)+v(z)-1,0\}}$, whenever these {\L}ukasiewicz operations can be defined.\\

\noindent Now, let us analyze the following assumption: Even before the verification, the lattice elements $x_i$ in the formula (\ref{2}) can be assigned truth values. Otherwise stated, it is conceivable that the propositions $X_i$ are in possession of pre-existing (i.e., existing before the detectors' clicks) truth values which are either merely revealed or somehow transformed by the verification.\\

\noindent In agreement with the analyzed assumption, let us suppose that before the verification the elements $x_1$ and $x_2$ are either the bottom element and the top element of a lattice or other way around. In such a case, prior to the verification, $v(\sqcup_{i=1}^{2} x_i ) = v(0_L \sqcup 1_L) = 1$ while $v(\sqcap_{i=1}^{2} x_i ) = v(0_L \sqcap 1_L) = 0$, and therefore, $v(x_{12})=1$ in accordance with the formula (\ref{2}).\\

\noindent Let $\mathbb{P}$ be the probability function mapping any proposition $Y,Z,\dots$ to the real interval $[0,1]$ such that $\mathbb{P}[Y]=1$ if $Y$ is true, $\mathbb{P}[Y]=0$ if $Y$ is false, and $\mathbb{P}[Y \lor Z]=P[Y]+P[Z]-P[Y \land Z]$. Along these lines, the probability function $\mathbb{P}$ can be considered as the degree of belief that the corresponding proposition is true (or it is expecting to be true).
$\,$\footnote{\label{f4}This approach to the generalization of the notion of a probability function allows to accommodate variation in the background logic of the account while maintaining the core of standard probability theory \cite{Weatherson}.\vspace{5pt}}\\

\noindent Since in the considered case ${[\![ X_1\lor X_2 ]\!]}_v =1$ and ${[\![ X_1\land X_2 ]\!]}_v =0$, the probability function mapping the conjunction $X_1 \lor X_2$ to the interval $[0,1]$ can be written by the sum of the probabilities $\mathbb{P}[X_1 \lor X_2 ] = \mathbb{P}[X_1] + \mathbb{P}[X_2] = 1$. So, were the pre-existing truth values of the propositions $X_1$ and $X_2$ to be such that $v(x_{12})=1$, the interference pattern $\mathbb{P}[R|X_1 \lor X_2]$ (i.e., the probability of finding the particle at a certain region $R$ on the screen) in the two-slit set-up with none of the detectors present at the slits would be the sum of one-slit patterns $\mathbb{P}[R|X_1]$ and $\mathbb{P}[R|X_2]$, namely, $\mathbb{P}[R|X_1 \lor X_2] =\text{\textonehalf}\mathbb{P}[R|X_1] + \text{\textonehalf}\mathbb{P}[R|X_2]$ (on condition that $\mathbb{P}[X_1] = \mathbb{P}[X_2]$), and thus the second-order interference term $I_{12} \equiv \mathbb{P}[R|X_1 \lor X_2] -\text{\textonehalf}\mathbb{P}[R|X_1] - \text{\textonehalf}\mathbb{P}[R|X_2]$ would be absent.
$\,$\footnote{\label{f5}The wording ``the second-order interference term'' is from \cite{Sinha}.\vspace{5pt}}\\

\noindent By contrast, let us suppose that before the verification $v(x_1)=1$ and $v(x_2)=1$ (which could be if both $x_i=1_L$) and so, according to the formula (\ref{2}), $v(x_{12})=0$ prior to the verification. But then, one would find – in contradiction to the quantum collapse postulate – that $v(\sqcap_{i=1}^{2} x_i ) = {[\![ X_1\land X_2 ]\!]}_v =1$, that is, it is not true that only one detector will click if the particle’s passage through the slits is observed.\\

\noindent Thus, the compound proposition $X_{12}$ would be in possession of pre-existing truth values (consistent with the occurrence of quantum interference and quantum collapse) only on condition that $v(x_i) \neq 0$ and $v(x_i) \neq 1$. Clearly, this condition could be met if prior to the verification, $X_1$ and $X_2$ did not obey the principle ``\textit{a proposition is either true or false}'', i.e., the principle of bivalence.\\

\section{Many-valued logics vs. supervaluationism}\label{Three}  

\noindent From the violation of bivalence, one can infer that results of future non-certain (not consistent with always true and always false propositions) events can be described using many-valued logics.\\

\noindent For example, in a series of papers \cite{Pykacz94,Pykacz95,Pykacz00,Pykacz10,Pykacz11,Pykacz15}, it is argued that for any lattice element $y \in L$ one should have\smallskip

\begin{equation} \label{4} 
   \left\{
      v\left(y\right) 
   \right\}
   =
   \left\{\mathfrak{v} \in \mathbb{R}\,|\, 0 \le \mathfrak{v} \le 1 \right\}
   \quad
   \text{whereas}
   \quad
   v(0_L)=0
   \;\,
   \text{and}
   \;\,
   v(1_L)=1
   \;\;\;\;  ,
\end{equation}
\smallskip

\noindent which implies that an infinite-valued logic should be used to describe not-yet-verified properties of quantum objects.\\

\noindent But what is more, from the violation of the principle of bivalence it is also possible to conclude that truth values of the future non-certain events simply do not exist, that is,\smallskip

\begin{equation} \label{5} 
   \left\{
   v\left(y\right) |  
   \, y \neq 0_L
   \;\: \text{and}
   \;\: y \neq 1_L
   \right\}
   =
   \emptyset
   \quad
   \text{whereas}
   \quad
   v(0_L)=0
   \;\,
   \text{and}
   \;\,
   v(1_L)=1
   \;\;\;\;  .
\end{equation}
\smallskip

\noindent Unlike the assumption of pre-existing many-valuedness (\ref{4}) which supposes that borderline (that is, uncertain) statements should be assigned truth-values lying anywhere between the truth and the falsehood, the assumption of \textit{supervaluationism} (\ref{5}) suggests that such statements should lack truth-values at all. This can neatly explain why it is impossible to know in advance the truth-values of the borderline propositions $X_1$ and $X_2$ concerning the path that the particle can take getting through the slits.\\

\section{Concluding remarks}\label{Four}  

\noindent Suppose a double-slit quantum interference experiment is described in the following manner: After the verification, the proposition that the particle passes through a particular slit comes out true. Now, let us ask the question, \textit{is this a complete description of the quantum interference experiment?}\\

\noindent The first answer is \textit{no}: In a complete description, the particle passes through either slit regardless of the verification since any specific proposition about the properties of the combined system (the particle + the detectors) can be not only either true or false but also \textit{neither true nor false}. Accordingly, in the complete description, all the elements of a lattice represent properties which the system can possess to some degree of truth.\\

\noindent The second answer is \textit{yes}: Prior to the detectors' clicks, the particle is by no means has passed through either slit. If both slits are opened, the passage through the given slit only comes about when the corresponding detector confirms it. As a result, the sentence ``\textit{the particle passes through a particular slit}'' can be a proposition, that is, a primary bearer of truth-value, only after the detectors have verified the particle's passage through the slit.
$\,$\footnote{\label{f6}One can easily notice that the description of a double-slit interference experiment presented above bears a great deal of similarity to Einstein’s example of a particle confined to a two-chambered box. See the detailed analysis of this example in \cite{Norton}.\vspace{5pt}}\\

\noindent It is clear that the assumption of pre-existing many-valuedness, specifically, infinite-valuedness, coincides with the first answer. Whereas the assumption of supervaluationism, as per which any element of a lattice other than the greatest element $1_L$ and the least element $0_L$ carries no truth values, corresponds to the second answer.\\

\section*{Acknowledgment}
\noindent The author would like to thank the anonymous referee for the inspiring feedback and the insights.\\

\bibliographystyle{References}
\bibliography{References}

\end{document}